\documentclass[12pt]{article}

\usepackage{amssymb} \usepackage{amstex} \usepackage{epsfig}
\newtheorem{conjecture}{Conjecture}
\newtheorem{theorem}{Theorem}

\begin{document}

\begin{titlepage}

\centering

{\bfseries \huge Detection of Aliasing in\\
                  Persistent Signals}

\bigskip

{\large 
{\bfseries 
Kevin R. Vixie\footnote{XCM MS-F645, Los Alamos National Laboratory, Los
Alamos, NM 87545, e-mail: vixie@@lanl.gov}
}

\smallskip

{\normalsize \itshape Los Alamos National Laboratory and

\small \itshape Portland State University}

\medskip

{\bfseries
David E. Sigeti\footnote{XCM MS-F645, Los Alamos National Laboratory, Los
Alamos, NM 87545, e-mail: sigeti@@lanl.gov}
\hspace{2.0 em}
Murray Wolinsky\footnote{TSA-5 MS-F602, Los Alamos National Laboratory,
Los Alamos, NM 87545, e-mail: murray@@lanl.gov}
}

\smallskip

{\normalsize \itshape Los Alamos National Laboratory}
}

\bigskip

{\large May 13, 1999}

\bigskip




\begin{abstract}
We explain why aliasing can be detected in a generic
temporally-sampled \emph{stationary} signal process.  We then
define a concept of stationarity that makes sense for single
waveforms. (This is done \emph{without} assuming that the
waveform is a sample path of some underlying stochastic process.)
We show how to use this concept to detect aliasing in sampled
waveforms.  The constraint that must be satisfied to make
detection of aliasing possible is shown to be fairly
unrestrictive.  We use simple harmonic signals to elucidate the
method.  We then demonstrate that the method works for
continuous-spectrum signals---specifically, for time series from
the Lorenz and R\"{o}ssler systems.  Finally we explain how the
method might permit the recovery of additional information about
Fourier components outside the Nyquist band.
\end{abstract}

\end{titlepage}

\newpage

\small{\tableofcontents}

\newpage

\section{Introduction}
\label{sec:intro}

Detection of aliasing from temporal samples alone, with no
restrictions on the original continuous-time source, is
impossible because any set of samples may be reconstructed (using
Shannon's sinc filter) to a properly sampled signal having the
same samples.  However, quite often additional information about
the source is available.  It is, of course, obvious that tight
constraints on the source would permit perfect reconstructions of
vastly undersampled signals.  For example, the constraint that
the data comes from a linear function of time makes any two
samples sufficient.  A less extreme example is a signal with a
lower as well as an upper frequency cutoff (a {\em bandpass
signal}).  For bandpass signals, it is well-known that one can
sample at a rate below twice the highest frequency while still
achieving perfect signal recovery (see~\cite[p.138, theorem
13.3]{higgins-1996}).

What are the weakest constraints that one can put on the signal
and still get something---detection of aliasing, for example?
Here, we examine constraints of stationarity.  In 1988 Hinich and
Wolinsky~\cite{hinich-1988} suggested a bispectral test for
detecting aliasing in temporally sampled stationary stochastic
processes\footnote{In the following, we will use the terms {\em
signal process} or just {\em process} for stochastic signal
processes.  Except when we use the terms {\em sample path} for a
realization of a stochastic process or {\em random sample}, the
word ``sample'' will refer to temporal sampling.}.  The test
aroused some controversy~\cite{{swami-1993}, {frazer-1993}} which
is examined in~\cite{hinich-1995,vsw1}.  In~\cite{vsw1} we show,
in detail, that the test {\em does} detect aliasing in some
signal processes and that it is the constraint of stationarity
that makes the detection of aliasing possible.  Briefly, if we
undersample a stationary process and then reconstruct a
continuous-time signal from the samples using the Shannon sinc
filter, the reconstructed process will not, in general, be
stationary.  In contrast, a {\em proper} sampling followed by
reconstruction will not destroy stationarity because this
procedure just reconstructs the original signal.  Detecting
nonstationarity in the reconstructed process thus suffices to
establish the existence of aliasing in the time series, provided
it can be assumed that the original signal was stationary.  These
results are reviewed in Section~\ref{sec:dasp}.

Applying these concepts requires either a random sample of the
paths of the process or an assumption of ergodicity which makes
it possible to extract statistics from a single sample path.  In
this paper we attempt to generalize the results for stationary
processes to the more common case where we have only a single
sample path and can make no assumption of ergodicity.  In other
words, we look for ways to discover undersampling in a time
series drawn from a single waveform, which may or may not be a
sample path of some underlying stochastic process.  We define
{\em sampling stationarity}, a form of stationarity that makes
sense for single waveforms, and show that it can be used to
detect aliasing in complex, continuous-spectrum signals.  We
present reasons to believe that sampling stationarity should be a
generic\footnote{We use the term \emph{generic} in a nontechnical
sense.  The term usually occurs in a situation where one would
like to say ``with probability $1$'' but where no obvious
probability measure exists.} property of signals and that the
destruction of sampling stationarity by the process of
undersampling and reconstruction should occur quite generally.
Finally, we explain how it might be possible to use the
reconstructed sample statistics plots (RSS plots) that we use to
detect aliasing to obtain additional information about individual
Fourier components beyond the Nyquist frequency.

The remainder of this paper proceeds as follows.  After
illustrating the key idea of this paper with an example in
Section~\ref{sec:intro_ex}, we proceed, in
Section~\ref{sec:dasp}, to demonstrate how a constraint of
stationarity permits the detection of undersampling in some
signal processes.  Then in Section~\ref{sec:dassp} we define
sampling stationarity.  In Section~\ref{sec:examples} we use
examples to show that the concept of sampling stationarity does,
indeed, enable detection of aliasing for nontrivial signals.  In
Section~\ref{sec:periodic} we consider the case of periodic
signals.  For this class of signals, we provide a complete
explanation of how (and when) the method of high-frequency
detection works.  The possible extension of this explanation to
nonperiodic signals is then discussed in
Section~\ref{sec:nonper}.  In Section~\ref{sec:RHFI}, we present
some reasons to believe that the plots that we have used to
detect aliasing may also be used to recover some portion of the
original signal's high-frequency content.  This is followed by
suggestions for further work (Section~\ref{sec:dfi}) and a
conclusion that summarizes the work in this paper
(Section~\ref{sec:concl}).  Two appendices contain computational
and mathematical details.

\section{Example}
\label{sec:intro_ex}

The key idea of our approach is captured by a very simple
example.  Suppose that we sample a square wave that takes the
values $-1$ and $1$.  There is a unique properly bandlimited
signal that has this time series as its samples.  We can compute
this signal by applying the Shannon sinc filter to our time
series.  We may regard this computation as an attempt to
reconstruct the original continuous-time signal.  If we can
reject this reconstructed signal as the source of our samples,
then we must conclude that the time series contains aliased
components.

Note that our given time series consists only of $-1$'s and
$1$'s.  The reconstructed signal, on the other hand, is
necessarily a continuous function of time, taking on all values
in the interval $[-1,1]$ (and, in fact, beyond).  The only way
that we could have obtained a sequence of $-1$'s and $1$'s by
sampling the reconstructed signal is if we had chosen a
particular sampling rate (or one of its subharmonics) and a
unique shift of the sampling comb.  Any other combination of
sampling rate and shift would have produced a series that takes a
continuum of values.  The probability of having chosen the
special sample rate and shift that give a sequence of $-1$'s and
$1$'s is clearly zero, provided that our sampling rate was chosen
independently of the source.  With this proviso, then, we can
reject (with probability $1$) the hypothesis that our time series
consisting of $-1$'s and $1$'s came from sampling the
reconstructed signal.

The assumption that the sampling rate was chosen independently of
the source is justified in most (but not all) cases of practical
importance because we can rule out any interdependence between
the source and the sampling rate on physical grounds.  For
example, if a signal produced by a distant source is sampled at a
predetermined rate, such a coupling is clearly out of the
question---it would amount to believing that the process that
produced the signal ``knew'' when we were going to sample at a
distant location.

We can draw valid conclusions from a sampled signal about Fourier
components beyond the Nyquist frequency only if we can put
constraints on the original continuous-time source.  How can we
characterize the constraints that we are imposing in this case?
Effectively, we are assuming that the sample times (which are
determined by the sample rate and shift) do not play a
distinguished role in the source.  Showing that the sampling
times are distinguished in the reconstructed signal then suffices
to reject the reconstructed signal as the original source of the
samples.

How, then, can we extend this analysis to more general classes of
signals?  In the case of a square wave (or any signal that takes
on a finite number of values), the appearance of the time series
produced by sampling the reconstructed signal at the given
sampling times could not be more different from the appearance of
a time series produced by sampling at any other shift of the
sampling comb.  Thus it is clear what we mean when we say that
the sample times are distinguished in the reconstructed signal.
For more general signals, however, it is not so clear exactly
what it means for the sample times to be distinguished.

There is one obvious case in which we can be assured that the
sample times are not distinguished in the original signal and in
which we can detect the distinguished character of the sample
times in the reconstructed signal.  If the original signal is, in
fact, a stationary signal process, then, by definition, {\em no}
time is distinguished.  The appearance of nonstationarity in the
reconstructed signal would then indicate the presence of aliasing
in the time series.  The detection of aliasing in time series
from stationary signal processes is the subject of the next
section.  Following that, we use our example of sampling from a
square wave and insights from the case of stationary processes to
develop a method for detection of aliasing in single waveforms.

\section{Detection of Aliasing in Stationary Processes}
\label{sec:dasp}

Consider the case of detection of aliasing in stationary signal
processes.  We start with the simplest stationary processes
imaginable---randomly shifted periodic signals.  If we have a
waveform, $x(t)$, with period $T$, then we can produce a
stationary process by adding to $t$ a random time shift,
$\theta$, that is evenly distributed on $[0,T)$.  A sample path
of our process then has the form $x(t+\theta)$ for a particular
choice of $\theta$.

Consider then the effect of undersampling and reconstruction on a
simple sine process,
\begin{equation}
    \label{eq:sin}
    x(t) = \sin(2 \pi f t + 2 \pi f \theta),
\end{equation}
where $\theta$ is evenly distributed on $[0,f^{-1})$.  If we
undersample with a sampling interval $\Delta t$, corresponding to
the Nyquist band $[-(2\Delta t)^{-1},(2\Delta t)^{-1})$, and then
reconstruct via convolution with the sinc filter, we get the sine
process given by
\begin{equation}
    \label{eq:sinrecon}
    x_r(t) = \sin(2 \pi \hat{f} t + 2 \pi f \theta).
\end{equation}
Here, $\hat{f}$ is the aliased frequency, given by $\hat{f} = f +
k_f/\Delta t$ where $k_f$ is the unique integer that places
$\hat{f}$ in the Nyquist band.  The key point is that the phase
of the reconstructed signal is the same as the phase of the
source even though the frequency has changed to the aliased value
$\hat{f}$.  For a process with a single harmonic, the
reconstructed signal remains stationary because the phase term,
$2 \pi f \theta$, is evenly distributed on $2\pi$.

Consider then a second signal process,
\begin{equation}
    \label{eq:twosin}
    x(t) = \sin(2\pi \alpha t + 2\pi\alpha\theta) +
           \sin(2\pi \beta t + 2\pi\beta\theta),
\end{equation}
where $\beta$ is an integer multiple of $\alpha$ and $\theta$ is
chosen randomly from the interval $[0,\alpha^{-1})$.  Since the
time shift, $\theta$, is the same for both components, this is,
for the various values of $\theta$, just a shifted waveform of a
given shape.  Since $\theta$ is evenly distributed over the
period, $\alpha^{-1}$, the process is stationary.

If we sample this process at a rate low enough for both
components to be aliased and then reconstruct using the sinc
filter, we get
\begin{equation}
    \label{eq:twosinrecon}
    x_r(t) = \sin(2\pi\hat{\alpha} t + 2\pi\alpha\theta) +
             \sin(2\pi\hat{\beta} t + 2\pi\beta\theta),
\end{equation}
where $\hat{\alpha}$ and $\hat{\beta}$ are the aliased
frequencies.  Although the phase terms, $2\pi\alpha\theta$ and
$2\pi\beta\theta$, are still evenly distributed over $2\pi$, they
now correspond to \emph{different} time shifts for the two
components.  Thus, we no longer have a single shifted waveform
and we can expect, in general, that stationarity will have been
lost.

\begin{figure}
\centerline{ \epsfxsize 2.5in \epsffile{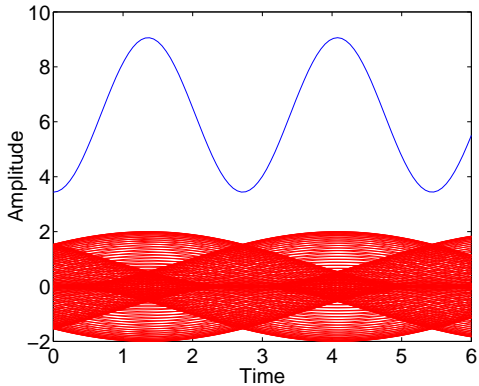}  }
\caption{Plot illustrating the nonstationarity of a randomly
shifted waveform that has been undersampled and then
reconstructed.  The upper curve is the sixth
moment\protect\footnotemark.  The lower curve shows the process
envelope.  The original process is given by
Equation~\ref{eq:twosin} with $\alpha=1.0$, $\beta=3.0$,
$\theta\in [0,1)$, and $\Delta t = e$.}
\label{fig:incomen}
\end{figure}

We illustrate this loss of stationarity on an example by setting
$\alpha = 1.0$ and $\beta = 3.0$ in Equation~\ref{eq:twosin} and
choosing a sample time, $\Delta t$, equal to $e$.  We may detect
the loss of stationarity by examining the envelope of the
reconstructed process.  We define the envelope of a process,
$X_t$, as the support of the probability density of $X_t$ as a
function of $t$.  For a process produced by randomly shifting a
periodic waveform, the envelope may be conveniently displayed by
plotting the sample paths corresponding to a representative
collection of time shifts as in Figure~\ref{fig:incomen}.
Clearly, a stationary process must have a constant envelope.  If
we compute the envelope for the process defined in
Equation~\ref{eq:twosinrecon} with the parameter values that we
have specified, we get an oscillating figure (see
Figure~\ref{fig:incomen}).  This implies that the signal is
nonstationary.  In fact, it is cyclostationary with period equal
to the sampling interval.

\protect\footnotetext{The sixth moment was chosen for
clarity of presentation.  The second moment remains constant in
this case.  The fourth moment does oscillate but the scale of its
oscillation is too small to allow meaningful display of the
moment and the envelope on the same scale.}

As explained above, we do not lose stationarity when our original
signal is a single sine wave.  Nor do we lose stationarity when
ratios between frequencies are preserved under the aliasing.  For
example, if $\Delta t = 1.0$ and the original frequencies are
$(10/9,20/9,30/9)$, they would alias to $(1/9,2/9,3/9)$ and we
would obtain another stationary process.  But this situation is
very special (non-generic).  In general, more than one Fourier
component is present and we do not have the special relationships
between the sampling rate and the component frequencies that
preserve ratios between frequencies when undersampling.  Thus, we
expect that, generically, a stationary process formed by randomly
shifting a periodic waveform will lose stationarity upon
undersampling and reconstruction.

\begin{figure}
\centerline{ \epsfxsize 2.5in \epsffile{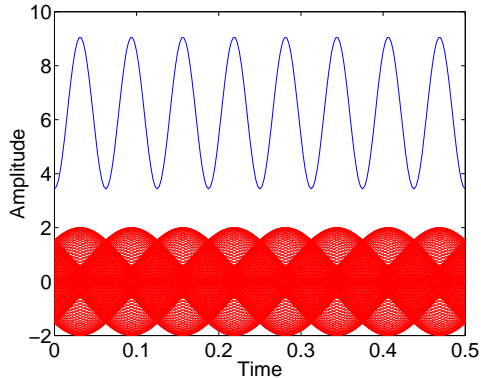}  }
\caption{The sixth moment (upper curve) and the process envelope
(lower curve) of the process given by Equation~\ref{eq:twosin}
with $\alpha=0.25$, $\beta=0.75$, $\theta \in [0,4)$, and $\Delta
t = 1.0$.}
\label{fig:comen}
\end{figure}

Within the context of single shifted waveforms, the destruction
of stationarity can occur in some remarkable situations.
Consider that undersampling and reconstruction can break
stationarity even when only one of the components ($\beta$, say)
is aliased and when $\beta$ aliases to $\alpha$ or $-\alpha$.  In
other words, stationarity can be broken even when the two
components, after undersampling, lie right on top of each other.
We can see this by choosing $\alpha = 0.25$, $\beta = 0.75$, and
$\Delta t = 1.0$ in Equation~\ref{eq:twosin}.  The envelope for
the reconstructed process is shown in Figure~\ref{fig:comen}
where the nonstationarity is apparent.  (Of course, we cannot
possibly detect this loss of stationarity by examining only a
single sample path, since the sample path will never be more than
a single sine wave of some amplitude and phase.)

Not all stationary processes are randomly shifted periodic
waveforms.  What can we say about more general stationary
processes?  It is clear that, if there exist \emph{no} phase
relationships between any of the Fourier components of the
process, then undersampling and reconstruction will not destroy
stationarity.  For a generic stationary process, though, we would
expect at least some sets of components to exhibit phase
relations.  In that case, we would expect stationarity to be
destroyed because it is difficult to imagine how the destruction
of stationarity associated with one set of components could
somehow be canceled out by the presence of other incommensurate
components.

This argument, together with the observation that there is simply
no reason to believe that stationarity should be preserved under
undersampling and reconstruction, suggests that the loss of
stationarity should be a general feature of stationary processes.

\section{Detection of Aliasing in Single Sample Paths}  
\label{sec:dassp}

The method that we have just used to detect aliasing in a sampled
stationary process requires complete knowledge of the
discrete-time process obtained by sampling the original
continuous-time source.  Usually, however, we have available to
us only a single sample path.  Therefore, we require a method for
detecting aliasing in a single waveform which may or may not be a
sample path of a stochastic signal process.

\begin{figure}[h]
\centerline{ \epsfxsize 2.5in \epsffile{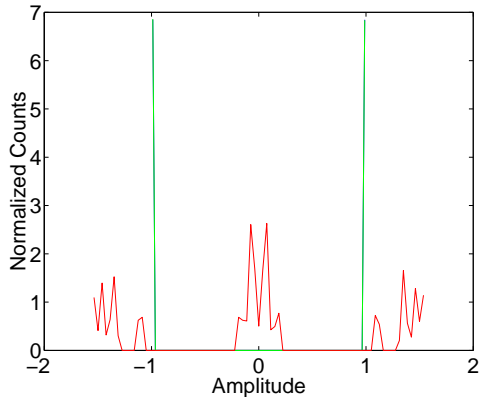}}
\caption{Sample statistics of data and reconstruction from a
square wave.  The plot shows sample statistics of the data in
blue and green (which are indistinguishable) and the
reconstruction in red.  (See the beginning of
Section~\ref{sec:examples} for an explanation of the blue and
green histograms.)  The red histogram is obviously very different
from the blue and green with which it would coincide if the
reconstruction had sampling stationarity.  The blue and green
histograms have been rescaled so as to make the three histograms
of comparable height.}
\label{fig:sq_stat}
\end{figure}

We may develop such a method by reconsidering the example of
sampling from a square wave discussed in
Section~\ref{sec:intro_ex} in light of our discussion of the
effect of undersampling on stationary signal processes.  Recall
that the sampled time series from the square wave takes on the
values $-1$ and $1$.  We may state this in statistical language
by saying that the one-time probability density of the time
series consists of two Dirac delta functions centered at $-1$ and
$1$, respectively.  Now, we would have obtained the same one-time
statistics if we had sampled the original square wave with any
shift of the sampling comb.  We will say that a waveform has
\emph{sampling stationarity} for a given sampling interval if the
one-time sample statistics do not change as the position of the
sample comb is shifted along the waveform.  \emph{Observing that
the original square wave had sampling stationarity for the given
sampling interval is essentially equivalent to saying that the
sample times were not distinguished in the source.}\footnote{Note
that the original square wave does not have sampling stationarity
for a sampling interval equal to its period.  In general, a
periodic signal will not have sampling stationarity with respect
to sampling intervals commensurate with its period.  However, the
set of sampling intervals that are commensurate with a given
period has Lebesgue measure zero.  Clearly, the probability of
choosing such a special sampling interval is $0$ under the
assumption that the sample times are chosen independently of the
source.}  Note that the signal obtained by applying the Shannon
sinc filter to the time series does \emph{not} have sampling
stationarity---the one-time statistics of the reconstructed
signal vary dramatically with shifts of the sampling comb (see
Figure~\ref{fig:sq_stat}).  This lack of sampling stationarity
corresponds to the distinguished role of the sample times in the
reconstructed signal.  Of course, this distinguished role for the
sample times is what allowed us to reject the reconstructed
signal as a candidate for the original source of the samples and,
thus, to conclude that the sampled series contained aliased
components.

This discussion suggests the following test for aliasing in
signals (no underlying stochastic process assumed).  Collect the
statistics on the recorded samples.  Reconstruct the signal at
various shifts of the sampling comb and collect the statistics at
these reconstructed samples.  Compare with the original
statistics.  (We use the term ``statistics'' loosely, without the
assumption that the samples are independent samples of some
underlying probability distribution.)  If we find that the
reconstruction has different statistics at some shift of the
sampling comb, an assumption of sampling stationarity for the
original signal implies that the reconstruction is not the
original signal and therefore that the signal was undersampled.

For this test to be at all useful, two questions must be
answered:
\begin{enumerate}
\item Are typical signals characterized by sampling stationarity?
      \label{question-1}

\item Do typical undersamplings reconstruct to signals for which
      sampling stationarity is violated? 
      \label{question-2}
\end{enumerate}

The answer to question~\ref{question-1} is clearly ``yes'' for
sample paths of ergodic stationary processes and for signals from
ergodic dynamical systems.  It is also clear that there are other
classes of signals which possess sampling stationarity.  For
example, general periodic signals (not just square waves) possess
sampling stationarity if the sampling interval is incommensurate
with the signal period (a generic condition).  Below, we
conjecture that sampling stationarity is a generic property of
signals.

The examples that we consider next suggest that the answer to the
second questions is also ``yes''.

\subsection{Examples}
\label{sec:examples}

In each of the examples listed below, the time series to which we
apply our test for aliasing was split into two interleaving
series, $D_1$ from the samples taken at [0,2$\Delta t$,4$\Delta
t$, ...] and $D_2$ from the samples taken at $[\Delta t,3\Delta
t,5\Delta t, ...]$.  The sample statistics corresponding to $D_1$
and $D_2$ are plotted in blue and green, respectively.  For
original signals with sampling stationarity, these two histograms
will coincide.  We then produce a reconstruction from $D_1$,
computed at the times corresponding to $D_2$.  The sample
statistics corresponding to this reconstructed series are shown
in red.  If the red histogram is significantly different from the
blue, then the reconstructed signal does not have sampling
stationarity.

\begin{figure}
\centerline{ \epsfxsize 2.5in \epsffile{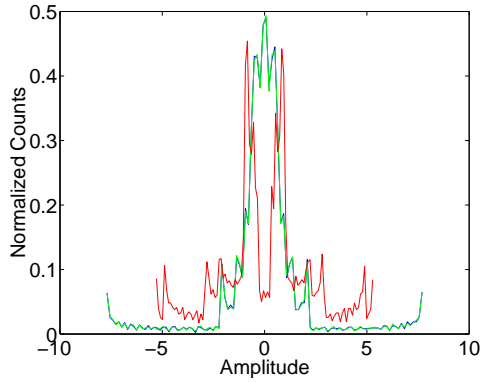}}
\caption{Sample statistics of data and reconstruction from a
periodic signal.  The blue and green histograms coincide,
indicating that the original signal had sampling stationarity.
The red histogram, showing the statistics of the reconstructed
signal, is obviously very different from the blue, showing that
the reconstruction does not have sampling stationarity.}
\label{fig:sines}
\end{figure}

{\bf Example:} For a periodic signal, the generic condition of
incommensurability of the sampling interval and the signal period
implies that the signal has the property of sampling
stationarity.  But we also find that undersampling and
reconstructing produces a signal that does NOT have sampling
stationarity, as illustrated in Figure~\ref{fig:sines}.  The data
for the plot were generated by sampling ($\Delta t = e/16$) a sum
of sines with frequencies $(0,1,2,...,10)$ and random amplitudes
that ranged between .78 and 1.22.

\begin{figure}
\centerline{ \epsfxsize 2.5in \epsffile{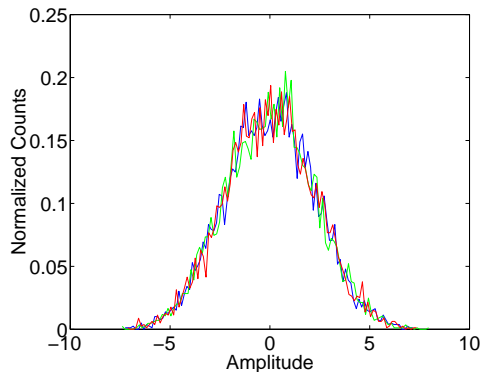}}
\caption{Sample statistics of data and reconstruction from a sum
of sine waves with incommensurate frequencies.  The blue and
green histograms coincide.  The red, showing the statistics of
the reconstructed signal, is NOT obviously different.}
\label{fig:quasi_sum}
\end{figure}

{\bf Example:} If our signal consists of a sum of sine waves with
incommensurate frequencies, then we cannot detect aliasing by
this method (see Figure~\ref{fig:quasi_sum}).  Although such a
signal will have sampling stationarity, the sampling stationarity
will not be broken by undersampling and reconstruction because it
is impossible to have relationships between the phases of
different Fourier components (see Section~\ref{sec:periodic}).

\begin{figure}
\centerline{ \epsfxsize 2.5in \epsffile{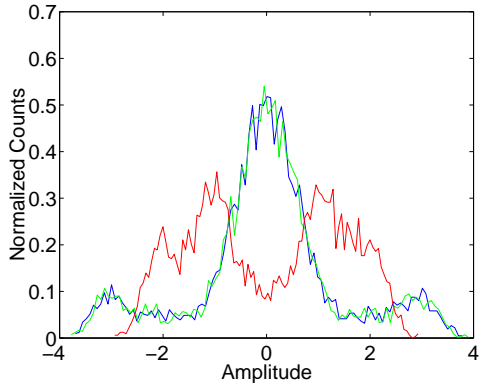}}
\caption{Sample statistics of data and reconstruction from a
mixture of a periodic waveform and incommensurate harmonics.
The blue and green histograms coincide.  The red, showing the
statistics of the reconstructed signal, is obviously different.}
\label{fig:mixed_sum}
\end{figure}

{\bf Example:} The previous example might lead to the suspicion
that this method works only for periodic signals (or step signals
such as the square wave).  However, the presence of
incommensurate Fourier components does not necessarily destroy
the ability to detect aliasing in a periodic waveform with more
than one harmonic component.  Figure~\ref{fig:mixed_sum} shows
the result of combining a periodic waveform with incommensurate
harmonics.  The total power in the incommensurate harmonics is
about $21\%$ of the power in the periodic waveform.  The sampling
stationarity of the original signal and the breakdown of sampling
stationarity with undersampling and reconstruction are apparent.
This shows that, as long as \emph{some} of our aliased Fourier
components are commensurate with other components, the method can
work.

\begin{figure}
\centerline{ \epsfxsize 2.5in \epsffile{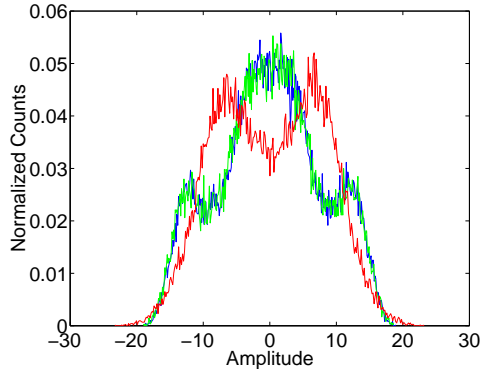} }
\caption{Sample statistics of data and reconstruction from the
$x$-coordinate of the Lorenz model.  The blue and green
histograms coincide and the red is clearly different.}
\label{fig:ergodic1}
\end{figure}

\begin{figure}
\centerline{ \epsfxsize 2.5in \epsffile{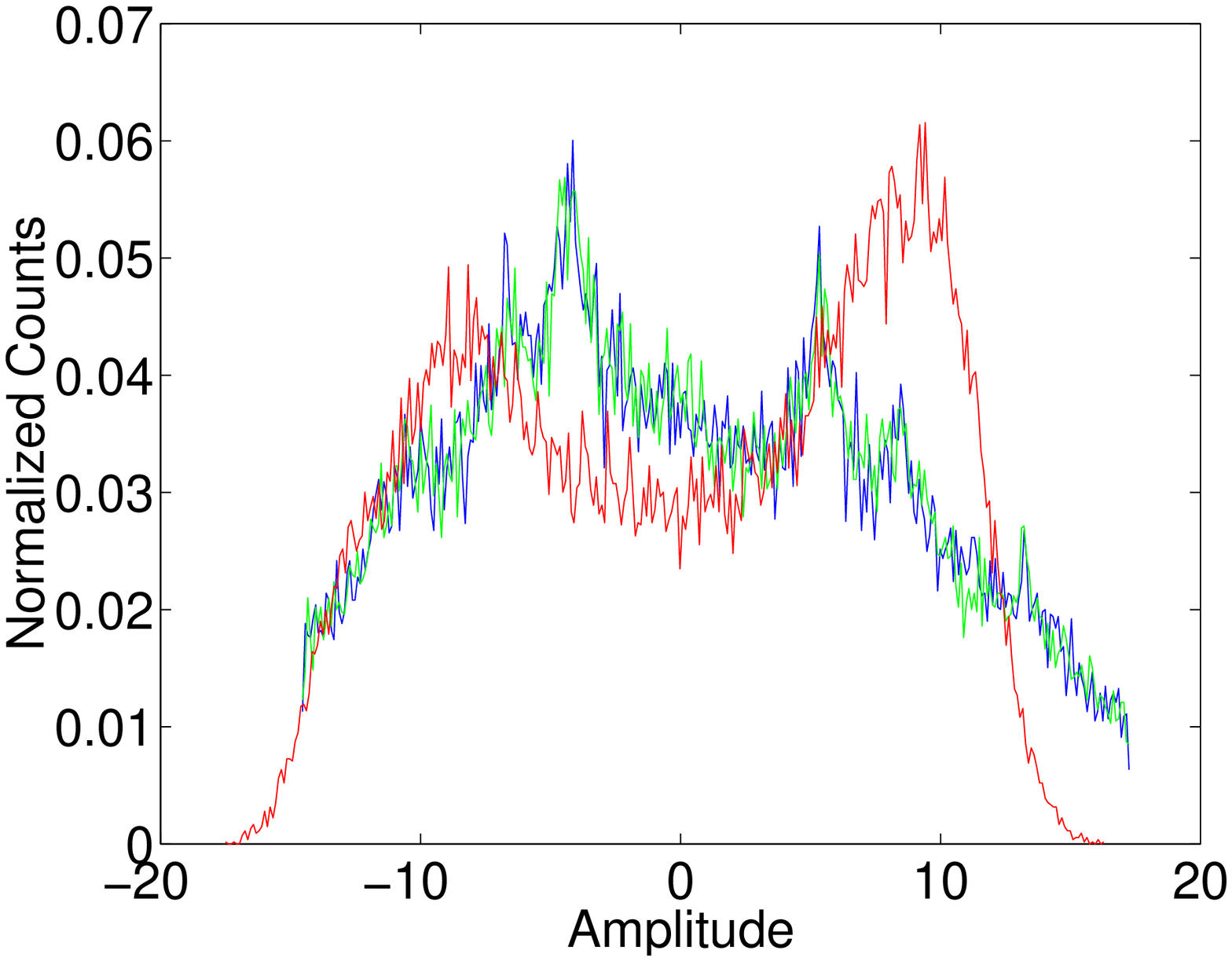} }
\caption{Sample statistics of data and reconstruction from the
$x$-coordinate of the R\"{o}ssler model.  The blue and green
histograms coincide and the red is clearly different.}
\label{fig:ergodic2}
\end{figure}

\begin{figure}
\centerline{ \epsfxsize 2.5in \epsffile{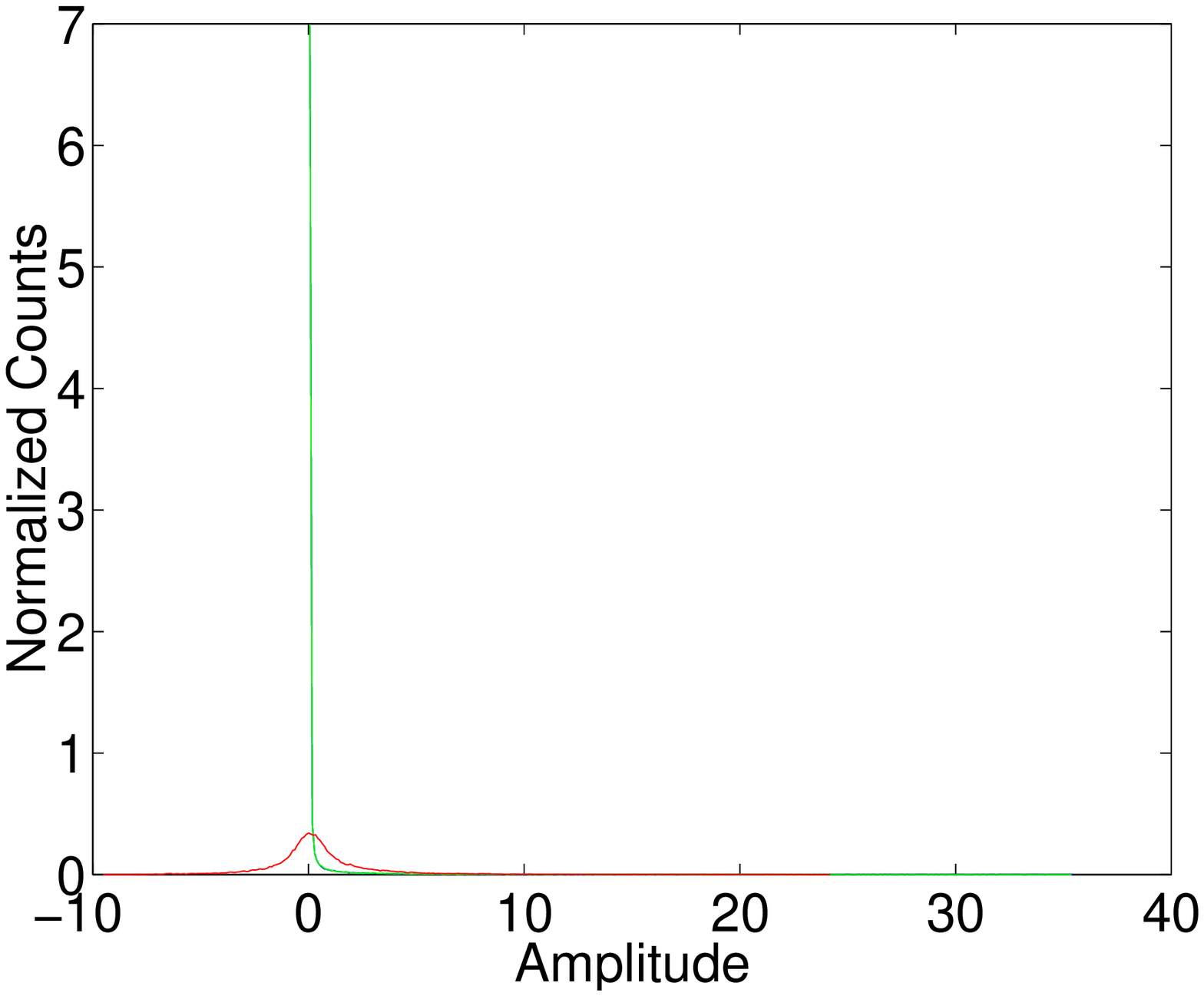} }
\caption{Sample statistics of data and reconstruction from the
$z$-coordinate of the R\"{o}ssler model.  The figure speaks for
itself.}
\label{fig:ergodic3}
\end{figure}

{\bf Example:} So far, we have demonstrated that the method works
for pure periodic signals and for periodic signals mixed with
incommensurate harmonics.  Figures~\ref{fig:ergodic1},
\ref{fig:ergodic2}, and~\ref{fig:ergodic3} show that the method
works for much more complex signals with continuous spectra.  The
signals are taken from the Lorenz and R\"{o}ssler systems (see
Appendix~\ref{sec:acd}).

The success in detecting aliasing in time series from the Lorenz
and R\"{o}ssler systems suggests that the method may work for a
very broad class of signals.  Before attempting to determine how
wide this class might actually be, we will look at the periodic
case in order to begin to understand the precise mechanism of the
method.

\subsection{A Closer Look at the Periodic Case}
\label{sec:periodic}

If one samples a periodic signal incommensurately with the signal
period, the samples end up mixing evenly around the waveform (see
Appendix~\ref{sec:append-waveform}).  Thus, all shifts of a
sampling comb with a sampling interval that is incommensurate
with the period will produce the same statistics.  This implies
that:
\begin{theorem}
    A periodic signal will have sampling stationarity with respect to
    any sampling interval that is incommensurate with the period of
    the signal.
\label{theo:periodic-incommensurate}
\end{theorem}
Conversely, sampling with an interval that \emph{is} commensurate
with the period will, in general, produce statistics that depend
on the sampling shift.
Theorem~\ref{theo:periodic-incommensurate} implies that:
\begin{theorem}
    Every periodic signal has sampling stationarity with respect
    to all sampling intervals except for a set of intervals with
    (Lebesgue) measure zero.
\label{theo:periodic-measure}
\end{theorem}
Thus, the probability of choosing a sampling interval for which a
given periodic signal does not have sampling stationarity is
zero, provided the interval is chosen independently of the signal.

What, then, is the effect of undersampling and reconstruction on
this sampling stationarity?  The sample statistics are determined
by the shape of the waveform (see
Appendix~\ref{sec:append-waveform} for the exact formula).  It
can be shown that the process of undersampling and reconstruction
is equivalent to sampling the original waveform at the same rate
with the individual Fourier components shifted with respect to
each other.  When different components experience different time
shifts, the shape of the effective waveform changes.
Consequently, the statistics change.  We now explain this in
detail.

When we sample a single harmonic with frequency $f$ and phase
$\varphi$ every $\Delta t$ time units, we get the values
\begin{equation}
  y_n = \sin\left(2\pi f n \Delta t + \varphi\right) \hspace{0.5cm} n \in Z.
\end{equation}
We will temporarily suppress the phase and rewrite this
expression as
\begin{equation}
\begin{split}
    \sin\left(2\pi f n \Delta t\right) 
        &= \sin\left(2\pi f n \Delta t + 2\pi k n\right)\\
        &= \sin\left(2\pi\left(f+\frac{k}{\Delta t}\right)n\Delta t\right)
\end{split}
\end{equation}
for any integer $k$, so that the reconstruction of this component
at points $1+s,2+s,3+s,...$ is given by
\begin{equation}
    \hat{y}_{n,s} 
    = \sin\left(2\pi\left(f+\frac{k_f}{\Delta t}\right)
                 \left(n+s\right)\Delta t\right),
\end{equation}
where the reconstruction chooses precisely one of the integral
$k$'s, which we will call $k_f$, such that $f+k_f/\Delta t$ is in
the interval $[-1/2\Delta t, 1/2\Delta t)$.  We can now rewrite
the reconstructed harmonic as
\begin{equation}
\begin{split}
    \hat{y}_{n,s} 
              &= \sin\left(2\pi\left(f+\frac{k_f}{\Delta t}\right) n \Delta t
                 + 2\pi\left(f+\frac{k_f} {\Delta t}\right) s\Delta t \right)\\
              &= \sin\left(2\pi f n \Delta t + 
                           2\pi k_f n + 2\pi f s \Delta t +  
                           2\pi k_f s\right)\\
              &= \sin\left(2\pi f n \Delta t + 
                           2\pi f s \Delta t + 2\pi k_f s\right)
\end{split}
\end{equation}
where we drop $2\pi k_fn$ since $k_f$ is an integer.  \emph{Thus,
the reconstructed signal has samples at a shift, $s$, as though
we were sampling the original waveform, but with the phase of the
individual Fourier component shifted by the amount $2\pi f s
\Delta t + 2\pi k_f s$.}  The first term amounts to a time shift
which is the same for all the components in the waveform.  This
implies that these first terms do not change the shape of the
waveform and can be ignored.  So we may consider the effective
waveform (at a shift $s$) to be
\begin{equation}
    \sum_i A_i \sin(2\pi f_i n \Delta t + 
                    2\pi k_{f_i}s + \varphi_i) 
\end{equation}
where we have reinserted the phase.  The term $2\pi k_{f_i}s$
amounts to a time shift that is different for different $f_i$.
This difference in time shifts leads to a change in the shape of
the effective waveform as $s$ changes which in turn changes the
sample statistics.

For a given waveform, it is clear that almost any change in the
shape of the waveform will change the sample statistics.  (For
example, a generic choice of $s$ will change the heights of the
extrema, changing the locations of the singularities in the
histogram.  See Appendix~\ref{sec:append-waveform}.)  Thus, we
conclude that \emph{generically, periodic signals have sampling
stationarity which is destroyed by undersampling and
reconstruction.}

\subsection{Nonperiodic Signals}
\label{sec:nonper}

Now, we want to use the insight that we have gained for the case
of periodic signals to get a better understanding of the answers
to the two questions at the end of Section~\ref{sec:dassp}.  We
begin with some general questions about the kinds of signals to
which our method might possibly apply.  

Consider first the case of transient signals.  In order to be
able to talk about sampling stationarity at all, we have to be
able to take as many samples as we want (at the given sampling
rate) in order to be able to estimate the one-time probability
distribution to arbitrary accuracy.  This implies that we must
think of our signals as functions of infinite time.  In this
context, any transient signal has trivial sampling
stationarity---the probability distribution is a delta function
at zero.  By the same token, undersampling and reconstruction
will not destroy this sampling stationarity.  Thus, we need to
restrict our attention to persistent (nontransient) signals.  

Within the class of persistent signals, it is clear that we need
the signals that we consider to have well-defined sample
statistics for arbitrary sampling intervals and shifts.  Given
that we are discussing aliasing, our signals also need to have a
Fourier transform (in some sense).  The set of signals with
well-defined power spectra (which will have, in general, singular
components) will clearly meet these criteria, although the actual
class to which our method applies may be larger.  In the
following, then, we may take the term \emph{persistent signal} to
refer to a signal with a well-defined, nonzero power spectrum.

Consider then the question of which signals have sampling
stationarity for which sampling intervals.  In the case of
periodic signals, Theorems~\ref{theo:periodic-incommensurate}
and~\ref{theo:periodic-measure} provide what is essentially a
complete answer---sampling stationarity holds for generic choices
of signals and sampling intervals.  At first glance one might try
to generalize Theorem~\ref{theo:periodic-incommensurate} to the
following:
\begin{conjecture}
    Every signal has the property of sampling stationarity for
    every sampling interval $\Delta t$ that is not commensurate
    with the period of any \emph{singular} component of its
    spectrum.  \emph{(FALSE)}
\end{conjecture}

Unfortunately this conjecture is false as may be seen from the
following counterexample.  If we undersample and reconstruct a
signal with a purely continuous spectrum (such as our signal from
the Lorenz system), we will introduce no new singular components.
Thus, the reconstructed signal will have a purely continuous
spectrum.  If the conjecture were true, then, such a
reconstructed signal would have sampling stationarity for all
sampling intervals by virtue of having no singular components.
Yet it is just the \emph{lack} of sampling stationarity of this
signal with respect to the given sampling interval that allows us
to detect aliasing in this case.  Thus, we know that there exist
signals that lack sampling stationarity with respect to sampling
intervals that are not commensurate with any singular component
of their spectra and the conjecture is false.  However, the
reconstruction of an undersampled signal has a very special
relationship to the interval with which the sampling was done.
Thus, one expects that resampling the reconstruction with a new
sampling interval \emph{not related to the original interval}
will yield statistics that are again stationary with respect to
shifts in the sampling comb.  Therefore, we arrive at the
following conjecture:
\begin{conjecture}
     Every signal has the property of sampling stationarity for
     \emph{every} sampling interval $\Delta$t, \emph{except} a
     set of $\Delta t$'s with (Lebesgue) measure zero.
\end{conjecture}

This conjecture implies that, if one were to observe the
reconstructed statistics varying with changes in the shift, this
observation would be enough to conclude (with probability $1$)
that the samples came from an undersampled waveform.  In other
words, \emph{the truth of the conjecture would imply that the
detection of undersampling by the proposed method is generically
free of false positives}

Next, we want to know when undersampling and reconstruction of
persistent nonperiodic signals will yield new signals which have
lost the property of sampling stationarity.  (In other words, we
also want to know when we can get false negatives.)  The analysis
that we have presented for periodic signals suggests that
undersampling and reconstruction should destroy sampling
stationarity for general persistent signals in which at least
some of the aliased Fourier components are commensurate with
other components of the signal.  The reasoning is that each
individual component can be regarded as a part of a family of
harmonics and that the effective shape of the waveform associated
with this family is changing with shifts of the sampling comb.
There does not appear to be any reason to believe that combining
different periodic waveforms, each of which is changing its
sample statistics with shifts of the sampling comb, would result
in sample statistics that do not change.  Note that the condition
that the original signal must have at least \emph{some}
commensurability will be satisfied by any signal with a nonzero
continuous part to its spectrum as well as by periodic signals.
Therefore, \emph{it seems likely that, generically, persistent
signals have sampling stationarity that is destroyed by
undersampling and reconstruction.}  In order to turn this last
statement into a well defined conjecture, it will be necessary to
define precisely what is meant by ``generically'' in the case of
persistent nonperiodic signals.  The question of exactly how to
define ``persistent'' must also be answered.  Since the
transformation that takes us from a waveform to sample statistics
is extremely nonlinear, a proof is likely to be difficult.

\subsection{Recovery of High-Frequency Information}
\label{sec:RHFI}

The next question that presents itself is whether or not we can
recover information about individual aliased Fourier components
using the sampling-shift dependence of the reconstructed
statistics.  Ideally, we would like to know how much of the
signal at an individual frequency, $f$, in the Nyquist band comes
from each frequency that aliases to $f$.

Consider the one-time probability density of the reconstructed
signal, $p(x)$, as a function of both $x$ and shift $s$.  We call
this two-dimensional surface a Reconstructed Sample Statistics
(RSS) plot.  The RSS plot has dependencies on $s$ tied directly
to the quantities $k_{f_i}$.  Each $k_{f_i}$, in turn, determines
the particular copy of the Nyquist band in which its
corresponding $f_i$ is located.  This chain of dependencies
suggests that the RSS plot contains the information necessary to
determine the contribution of each band to the signal at a given
frequency in the Nyquist band.  The inverse problem is greatly
complicated by the interaction of the Fourier components and the
nonlinear ``projection'' that turns the waveform into statistics.
This extremely nonlinear inverse problem is the main object of
our current research.


\section{Directions for Further Investigations}
\label{sec:dfi}

In addition to the work already alluded to on the inverse problem
formed by the RSS plots, there are other issues to explore.
Included among them are:

\begin{itemize}
\item What are the effects of noise on this method for detection
of aliasing?

\item What is the effect of near commensurability of sample
interval and signal period?

\item What is the effect of finite time-series length?

\item How does the departure of the statistics of the
reconstructed signal from stationarity depend on the fraction of
the total power that lies outside the Nyquist band?

\end{itemize}

These questions are important to the practical usefulness of the method
of high-frequency detection/recovery.

\section{Conclusion}
\label{sec:concl}

Although the idea of detection of aliasing is typically dismissed
with references to the Nyquist criterion and the Shannon
reconstruction theorem, we have demonstrated that detection of
aliasing is possible with what appear, at first glance, to be
very weak prior assumptions.  The key concept is that of sampling
stationarity.  We emphasize that \emph{this concept makes sense
for single waveforms.}  Although this concept arose in the
consideration of step signals like square waves, its usefulness
extends far beyond these signals.  In particular, our method
enables the detection of aliasing in samples from nontrivial
waveforms such as measurements from motion on the Lorenz or
R\"{o}ssler attractors.  As indicated above, many questions
remain.  Some of these are important for the practical utility of
the concept of sampling stationarity and the associated RSS
plots.

\appendix

\section{Computational Details}
\label{sec:acd}

The calculations represented in the paper were done with Matlab.
The Lorenz equations,
\begin{equation}
  \begin{split}
      \dot{x} =& \sigma(y-x) \\ \dot{y} =& x(R-z) - y  \\ \dot{z} =& x
      y - b z,
  \end{split}
\end{equation}
were integrated with parameter values of $\sigma = 10$, $R = 28$,
and $b = 8/3$ using Matlab's ``ODE45'' which is an adaptive step
size routine.  Relative tolerance was set to the default value of
$1.0 \times 10^{-3}$ and absolute tolerance was set to the
default $1.0 \times 10^{-6}$.  Initial conditions were set at
$x=y=z=1$.  Values for the $x$, $y$, and $z$ coordinates were
saved every 0.5 time units.  200,001 samples were taken and split
into two interleaving time series each 100,000 samples long.  The
first series was used to reconstruct a signal via convolution
with a sinc filter of length 200,001.  These very long series and
filters were used to minimize the effects of truncating the
convolution at the ends of the series.  The histograms for the
reconstructed signal were computed from the middle 50\% of the
reconstructed series.

The R\"{o}ssler equations,
\begin{equation}
  \begin{split}
      \dot{x} =& -z-y       \\ \dot{y} =& x + a y     \\ \dot{z} =& b
      +  z (x - c),
  \end{split}
\end{equation}
were integrated in the same way with a sampling interval of 10
time units, and parameter values of $a = 0.15$, $b = 0.2$, and $c =
10$.  In this way 200,001 samples were obtained and the splitting
and reconstruction were done as described for the Lorenz
equations.

All the histograms presented here were originally calculated from
much shorter time series ($10,000$ samples).  The features that
allow us to conclude that aliased components are present were all
clearly visible in the histograms made from shorter time series
although, of course, the histograms were considerably rougher.
We conclude that the results that we have presented are certainly
not an artifact of finite-length time series.

\section{Sample Statistics for a Periodic Signal}
\label{sec:append-waveform}

If one samples a periodic signal, $h(t)$, incommensurately with
the signal period $T$, the samples end up mixing evenly around
the waveform\footnote{Proving this is surprisingly difficult.  An
equivalent problem is the determination of the density of points
obtained by repeated iterations of an irrational rotation on a
circle of unit circumference.  The resulting distribution of
points satisfies $n_{[a,b)}/N \approx (b-a)$ where $n_{[a,b)}$ is
the number of iterates in $[a,b)$, N is the total number of
iterates, and $[a,b) \subset [0,1)$.  See
~\cite[p.39-40,29]{sinai-1976} for details.}.  The resulting
histogram is proportional to the reciprocal of the derivative of
the waveform.  This follows from the fact that the probability of
getting any particular $t$ (position along the waveform) is
uniformly distributed over $[0,T)$ which in turn implies that the
probability of the interval $[y,y+dy)$ is the probability of the
corresponding $dt$ or $(1/T)(dy / h^{'}(t))$.  More precisely,
the probability density for $y$ is
\begin{equation}
  \label{eq:project}
  p(y) = \frac{1}{T} 
         \sum_{t_\alpha \in \mathcal{T}(y)} 
             (h^{'}(t_\alpha))^{-1}
\end{equation}
where
\begin{equation}
        \mathcal{T}(y) = \{t \mid t \in [0,T), h(t) = y\}
\end{equation}

Note that the density will have $1/\sqrt(y)$ singularities at the
local maxima and minima of $h(t)$.  The form of the singularities
follows from the fact that a generic waveform has maxima and
minima with nonzero second derivative.

\bibliographystyle{plain}
\bibliography{the_bib}

\begin{thebibliography}{1}

\bibitem{frazer-1993}
G.~Frazer, A.~Reilly, and B.~Boashash.
\newblock The bispectral aliasing test.
\newblock In {\em Proceedings IEEE Workshop on Higher-Order Statistics}, pages
  332--335, June 1993.

\bibitem{hinich-1995}
Melvin~J. Hinich and Hagit Messer.
\newblock On the principal domain of the discrete bispectrum of a stationary
  signal.
\newblock {\em IEEE Transactions on Signal Processing}, 43(9):2130--2134, 1995.

\bibitem{hinich-1988}
Melvin~J. Hinich and Murray~A. Wolinsky.
\newblock A test for aliasing using bispectral analysis.
\newblock {\em Journal of the American Statistical Association}, 83:499--502,
  June 1988.

\bibitem{higgins-1996}
J.R.Higgins.
\newblock {\em Sampling Theory in Fourier and Signal Analysis}.
\newblock Oxford University Press, 1996.

\bibitem{sinai-1976}
Ya.~G. Sinai.
\newblock {\em Introduction to Ergodic Theory}.
\newblock Mathematical Notes. Princeton University Press, 1976.

\bibitem{swami-1993}
A.~Swami.
\newblock Pitfalls in polyspectra.
\newblock In {\em Proceedings of the IEEE International Conference on
  Acoustics, Speech and Signal Processing}, volume~IV, pages 97--100, 1993.

\bibitem{vsw1}
Kevin~R. Vixie, David~E. Sigeti, and Murray Wolinsky.
\newblock On the bispectral aliasing test.
\newblock {\em in preparation}, 1999.

\end{thebibliography}

\end{document}